# Comparison of residual dose rates from MARS and DORIAN simulations with experimental data from the SINBAD database

A. Makovec*, I. Tropin
Fermi National Accelerator Laboratory, Batavia, Illinois 60510, USA
*amakovec@fnal.gov

*Abstract*

This benchmarking study aims to compare residual dose rate predictions from the MARS and DORIAN codes with experimental measurements taken from samples irradiated at CERN's CERF facility. The research evaluates the residual dose rates of these samples over varied cooling intervals. Additionally, it delves into the influence of parameter choices, such as specific physics processes and energy thresholds, on the production and transport of particles and their effects on simulation outcomes. While both MARS and DORIAN largely align with experimental findings, discrepancies were observed in the copper sample. This prompted an in-depth examination of the elevated dose rates from MARS.

## INTRODUCTION

Computer codes and nuclear data play a crucial role in predicting and understanding the behavior of ionizing radiation in various shielding scenarios. The accuracy and reliability of these predictions are important for various fields such as high energy physics, nuclear energy, and medical applications. To ensure the validity and robustness of computer codes and nuclear data, it is crucial to validate and benchmark them against experimental data.

The SINBAD database (an International Database for Integral Shielding Experiments) provides a large set of experimental data on radiation shielding and transport problems [1]. The goal of the SINBAD database is to preserve a unique set of experiments for the needs of today and tomorrow, and to provide a reference for the validation and benchmarking of computer codes and nuclear data.

In this paper, we selected the study called "BENCHMARK STUDY OF RESIDUAL DOSE RATES WITH FLUKA" [2] from the SINBAD database as experimental dataset to compare the results obtained from the Monte Carlo code MARS [3, 4, 5] and the DORIAN code [6], which is an extension of the FLUKA Monte Carlo simulation code [7, 8]. The aim of this study is to compare the residual dose rates calculated by MARS and DORIAN with the experimental data from the SINBAD database, and to assess the accuracy and reliability of the codes for activated samples.

## METHODOLOGY OF THE SINBAD BENCHMARK

In the CERN-EU high-energy Reference Field (CERF) facility [9], samples made of materials commonly used in accelerator machines and shielding were exposed to a pulsed, 120 GeV/c mixed hadron beam (comprising 1/3 protons and 2/3 positively charged pions) from the Super Proton Synchrotron (SPS), which was impinged on a copper target.

Following the irradiation of the samples, the residual dose rate was assessed at varying cooling times ranging from approximately 20 minutes to one month. The measurements were conducted both at contact and at distances of 12.4 cm, 22.4 cm, and 32.4 cm, calculated from the surface of the sample – that faced the CERF target during irradiation – to the center of the detector. A portable spectrometer was utilized for the measurements, performed in a laboratory with a low background radiation dose rate of 55 nSv/h. The spectrometer was based on a cylindrical NaI crystal with a diameter and height of approximately 5 cm.

The SINBAD database documentation highlights three primary sources of uncertainty in the measurements. First, a 2 mm uncertainty exists in determining the detector's effective center. Second, the positioning of the sample in its holder introduces another 2 mm uncertainty regarding its distance from the detector. Lastly, a systematic instrument error of 1 nSv/h is associated with the displayed readings.

Given these geometrical uncertainties and employing the inverse square law, we can deduce the bounds of uncertainty for the dose rate measurements at different distances. A 4 mm offset was applied in most cases, except for the contact measurement where physical constraints allowed only a 2 mm offset. At contact, the geometry-related uncertainty ranged from -25.7% to 18.1%. In comparison, at 30 cm, the uncertainty was between -2.6% and 2.7%. Due to the higher uncertainties at contact, it's advisable to emphasize comparisons between simulated and experimental findings at greater distances. Additionally, a systematic instrument uncertainty of ±1 nSv/h should be considered for all measurements.

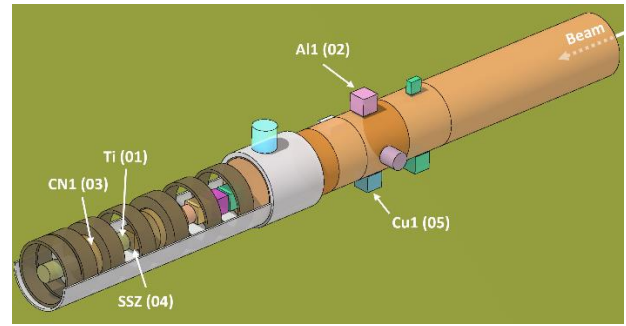

*Figure 1: 3D model of the experimental setup layout from FLUKA.*

As part of the benchmark study conducted within the SINBAD database, the FLUKA code was utilized to simulate the CERF irradiation experiment. Distinct simulations were performed for both proton and pion beams, and the results were merged in a post-processing stage to comply with the actual beam composition. Simulation of the electromagnetic cascade generated by gamma and positron emitters was carried out for each cooling time, however, the emission of electrons was disregarded as it was found to have a negligible contribution to the total dose rate. The equivalent dose was determined by multiplying the particle fluence with the relevant fluence-to-dose equivalent conversion factors.

Simulation parameters: The standard FLUKA preset, "NEW-DEFAULTS," dictated the physics processes. During the irradiation simulation, transport of electrons, positrons, gammas, muons, and neutrinos was omitted. An energy transport cut-off was set at 1 keV for all particles, except for neutrons and K_L, which had a threshold of 50 MeV. Full transport was applied to both light and heavy ions. The coalescence mechanism wasn't applied, and no superposition model (i.e., ion splitting into nucleons during nonelastic interactions) was in place. Additionally, photonuclear interactions and muon pair production by photons were disregarded. Pair production and bremsstrahlung by high-energy muons, charged hadrons, and light ions were also excluded from the simulation.

## RESIDUAL DOSE RATE CALCULATIONS WITH THE DORIAN CODE

The DORIAN code, which is written in the Python programming language, and based on the FLUKA Monte Carlo simulation code, was used to calculate residual dose rates in a three-step process, allowing for a comprehensive analysis of the various contributions. With the ability to change the geometrical configuration after the irradiation and quickly recalculate the dose rate for any stepwise irradiation profile and cooling time, DORIAN is an effective tool for optimizing residual dose rates in accelerator facilities. For this study, we applied the experimental parameters detailed in the SINBAD benchmark to the DORIAN calculations. This approach allowed us to directly benchmark our results against the experimental data, rather than comparing them with the FLUKA results documented in the SINBAD benchmark. The results presented in this paper are based on simulations performed using CERN FLUKA version 4-3.2 and DORIAN build July 2020, which utilizes Python 3.

### First Step

In the first step of the calculation, the FLUKA input file provided by the SINBAD benchmark was employed to simulate the primary process of interest. The physics processes and particle transport parameters were adjusted to match the experiment's specifics. We utilized the FLUKA preset named PRECISIOn, with the EMF option activated and without suppressing any particle transport. The energy thresholds for particle transport were set as follows: 1E-5 eV for neutrons, 100 keV for electrons/positrons (e-/e+), and 80 keV for photons. The kinetic energy production thresholds were likewise set at 100 keV for e-/e+ and 80 keV for photons. Full transport was used for both light and heavy ions. The Coalescence mechanism was activated, as was the Superposition model, in the range from 0.005 to 0.150 GeV/n. Photonuclear interactions, muon pair production, and muon nuclear interactions were all enabled. To increase the probability of photonuclear interaction, biasing of the hadronic inelastic interaction length was applied to photons and muons with a bias factor of 0.05. Additionally, pair production and bremsstrahlung by high-energy muons, charged hadrons, and light ions were considered.

The creation of radionuclides in the copper, iron, titanium, aluminum, and concrete samples (sample indexes 05, 04, 01, 02, and 03, respectively, as listed in Table 1 of the SINBAD benchmark) was recorded through the use of dedicated user routines (ftelos.f and usrrnc.f). Figure 1 shows the position of these samples during irradiation. Two separate simulations were run, one for protons and one for positively charged pions, with 2.5e5 and 5.0e5 primaries, respectively.

### Second Step

In the second step, FLUKA simulations were performed to compute the dose-per-decay contributions from the decay of the radionuclides produced in the first step. A simplified geometry was used, where the sample of interest was positioned in air and all other regions were removed. The DORIAN SOURCE user routine was utilized to read the recorded isotope information from the first step and to generate radioactive decay events at the recorded positions of each radionuclide. This resulted in the generation of decay positions for both the radionuclides themselves and all those in the decay chains originating from them. The decay of radioactive isotopes was enabled through the FLUKA RADDECAY card with the semi-analogue decay option and the energy thresholds for transport were set to 100 keV for e-/e+, and 80 keV for photons for the decay part of the simulation. The number of replicas of the decay of each individual residual was set to 30.

### Third Step

In the final step of the DORIAN process, the dose rate is calculated by combining the contributions from the various radionuclides determined in the previous step, based on the specified irradiation profile and cooling time. A separate DORIAN Python script was created for each sample to accommodate the various irradiation profiles listed in the SINBAD benchmark (Table 1). All calculations were based on a typical SPS cycle with a cycle length of 16.8 seconds and a 5.1-second spill at the beginning of each cycle. Cooling times were considered in the range of 0.1 to 1000.0 hours.

To determine the statistical uncertainty of the dose rate, a simple algorithm was developed and employed. This involved dividing the recorded isotope list into five parts, performing 5 parallel cycles of DORIAN's second and third step calculations using these parts as inputs, and then calculating the mean values and statistical uncertainties based on the results. Additionally, a removal tool was used to eliminate files that caused FLUKA crashes, which would occur with isotopes that had high half-life values (>10^9 seconds). Examples of such isotopes include Cr-50 ($T_{1/2}$= 1.8e17 y), Ca-48 ($T_{1/2}$= 5.3e19 y), and Pd-110 ($T_{1/2}$= 6.0e17 y).

## RESIDUAL DOSE RATE CALCULATONS WITH THE MARS CODE

The MARS code system is a collection of Monte Carlo programs that allow for detailed simulation of hadronic and electromagnetic cascades, muon, heavy-ion, and low-energy neutron transport in three-dimensional geometries that include shielding, accelerator, detector, and spacecraft components. It has the capability to simulate particles with energies ranging from a fraction of an electron volt to 100 TeV.

To ensure that the results of the simulation accurately reflected the conditions in the experiment being modeled, the beam parameters, irradiation profiles, material compositions, and geometries reported in the SINBAD benchmark were used as inputs to the MARS simulation. The 3D geometry was reconstructed using the ROOT geometry package (TGeo) to provide a precise representation for the MARS simulations.

In this paper, the MARS simulation incorporates the FLUKA simulation parameters from the SINBAD benchmark study, along with default settings for any parameters not detailed in the study. This approach not only enabled us to benchmark our results against experimental data but also allowed for a comparison of our simulation results with the FLUKA findings detailed in the SINBAD benchmark.

It is worth noting that at present, MARS does not handle nuclear isomers. In instances where there is the potential for the generation of an atomic nucleus in an energetically excited nuclear state, MARS instead registers a nucleus with the lowest-energy ground state.

## COMPARISON OF THE CALCULATED DOSE RATES WITH EXPERIMENTAL DATA

Table 1 lists the irradiation profiles applied to the copper, iron, titanium, aluminum, and concrete samples. Two samples were positioned laterally to the copper target (samples 05 and 02) and the other sample were placed downstream to the target, Figure 1. The number of beam particles on target was consistent across all samples, except for the concrete target, which had fewer beam particles despite merging four short irradiation profiles. Notably, the reported mass of the copper target in the SINBAD benchmark (28.253 g) was 21.4% less than the mass calculated from the available geometrical and density data (35.952 g). Based on our findings, however, it is likely that this discrepancy is due to a typo in the SINBAD paper.

*Table 1: Irradiation and sample parameters*

| Parameter\Sample | Cu1 (05) | SSZ (04) | Ti (01) | Al1 (02) | CN1 (03) |
|---|---|---|---|---|---|
| Irradiation profile | long 1 irrcyc_Aug 03_15.inp | long 4 irrcyc_Aug 03_13.inp | long 5 irrcyc_Aug 03_12.inp | long 2 irrcyc_Aug 03_16.inp | short 1-4 irrcyc_Aug 03_03.inp |
| No. of SPS cycles | 31133 | 29675 | 28412 | 33210 | 12812 |
| Total No. of beam particles | 1.43E+12 | 1.32E+12 | 1.26E+12 | 1.55E+12 | 3.88E+11 |
| Sample volume [cm$^3$] | 4.0 | 6.8 | 6.283 | 6.283 | 62.21 |
| Density [g/cm$^3$] | 8.988 | 6.676 | 4.42 | 2.718 | 1.7 |
| Mass (SINBAD) [g] | **28.253** | 45.378 | 27.796 | 17.055 | 102.853 |
| Mass (DORIAN) [g] | **35.952** | 45.397 | 27.772 | 17.078 | 105.759 |
| Sample position | lateral | downstr. | downstr. | lateral | downstr. |

In Figure 2 to Figure 6, the measured residual dose rates from the SINBAD benchmark are displayed along with the results of the "single step" MARS simulation and the DORIAN calculations for various samples. These graphs present the results for both contact (d=0) and for measurements taken at 30 cm (d=30 cm). The graphs' lower sections show the relative difference between the DORIAN calculations and the corresponding experimental measurements at both specified distances. It is evident from the graphs that there is good agreement between the DORIAN and SINBAD results and that there is generally a good agreement between the MARS results and the experimental data, except for the copper sample (05).

Figure 2 shows that for cooling times below approximately 100 hours, MARS results indicate higher dose rates from the copper sample in comparison to the experimental data. At contact, DORIAN's results align with the experimental data, once uncertainties are factored in. Conversely, at 30 cm, DORIAN's predictions fall within ±20% of the experimental figures.

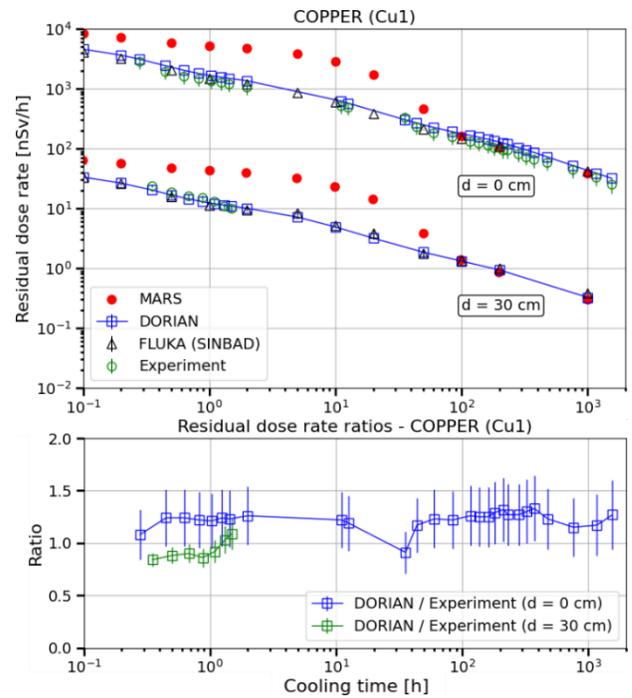

*Figure 2: Comparison of residual dose rates for the Cu1 sample (05)*

The residual dose analysis from DORIAN indicates that for the copper sample, the dose rate is dominated by $^{62}$Cu for cooling times less than 30 minutes, by $^{61}$Cu for cooling times ranging from 30 minutes to 2 hours, by $^{64}$Cu for cooling times 5 to 10 hours, by $^{52}$Mn for cooling times between 20 and 200 hours and by $^{58}$Co for 1000 hours of cooling.

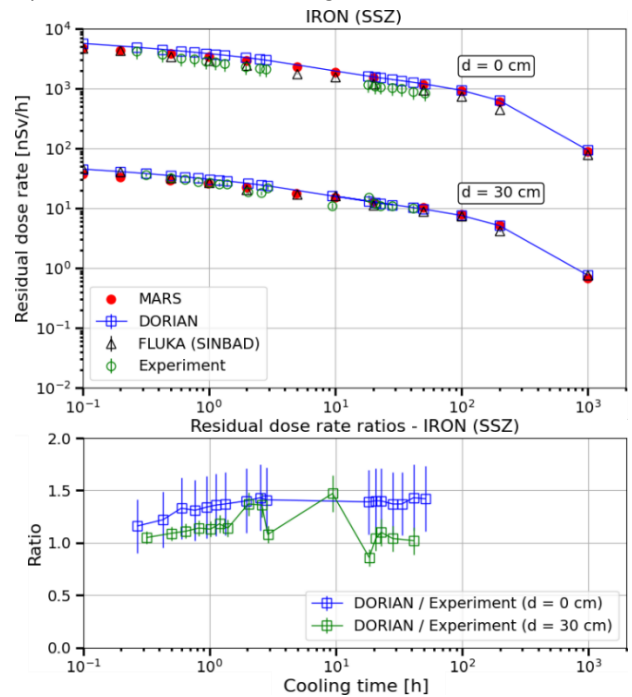

*Figure 3: Comparison of residual dose rates for the SSZ sample (04)*

The dose rate results for the iron sample are compared in Figure 3. Both the MARS and DORIAN results are in good agreement with the experimental data. However, it's noteworthy that both models yield slightly higher dose rates at contact. Interestingly, the measurements taken at 30 cm and at cooling times from 3 to 20 h exhibit unexpected fluctuations: a lower dose rate for shorter cooling times and vice versa. In case of the activated iron sample, for cooling times between 6 minutes and 200 hours, $^{52}$Mn is the primary contributor to the dose rate, with notable contributions from $^{44}$Sc between 6 minutes and 3 hours and from $^{48}$V between 20 and 200 hours. At a cooling time of 1000 hours, $^{48}$V becomes the dominant contributor to the dose rate.

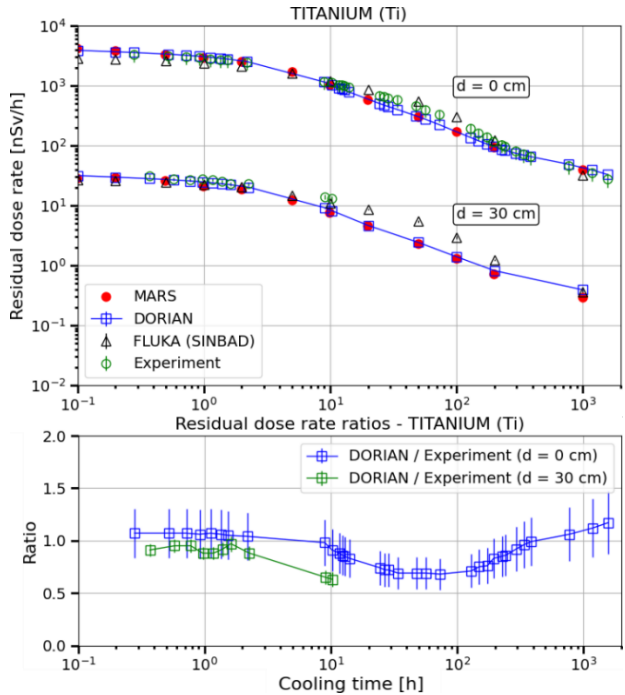

Figure 4: Comparison of residual dose rates for the Ti sample (01)

As seen in Figure 4, the dose rate results for the titanium sample from both MARS and DORIAN show good agreement with the experimental data, except for cooling times between 20 and 200 hours where both methods give slightly lower values than measured at contact. The dose rates for this range are dominated by $^{48}$Sc and $^{46}$Sc. For cooling times ranging from 6 minutes to 10 hours, $^{44}$Sc is the primary contributor to the dose rate, while for cooling times of 200 and 1000 hours, most of the dose comes from the decay of $^{46}$Sc.

Although experimental data is not available in this range for the 30 cm distance, it is worth noting that the observed discrepancy between the dose rates produced by DORIAN and the experimental data is consistent for the range of cooling times between 20 and 200 hours. Comparing the results for 10 cm and 20 cm, DORIAN gives slightly lower dose rates for cooling times between 20 and 200, and between 20 and 50 hours, respectively.

Figure 5 shows that for the aluminum sample, both MARS and DORIAN results are in good agreement with the experimental data, considering the uncertainties. DORIAN slightly overestimates dose rates at contact for cooling times under 200 hours. However, the dose rate predictions from the FLUKA simulation in the SINBAD benchmark also exhibit higher values in this range compared to experimental data, but to a lesser extent.

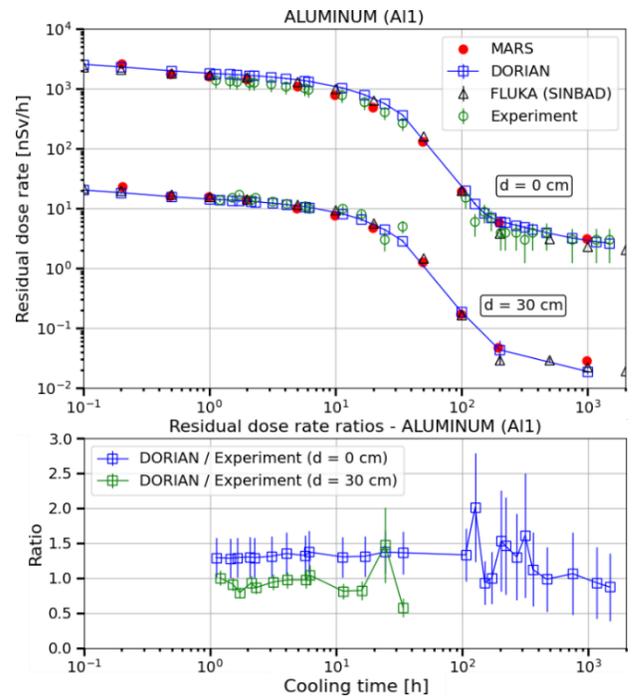

Figure 5: Comparison of residual dose rates for the Al1 sample (02)

The fluctuations observed in the experimental data within the ratio plot between 100 to 400 h are consistent with the high geometrical uncertainties at contact. At 30 cm, DORIAN shows good agreement with the experimental data, except for the last 2 values: these, however, show unexpected fluctuations (lower dose rate for shorter cooling time). For the Aluminum sample the dose rate is dominated by $^{24}$Na for cooling times ranging from 6 minutes to 100 hours, and by $^{48}$V and $^{22}$Na for cooling times of 200 and 1000 hours.

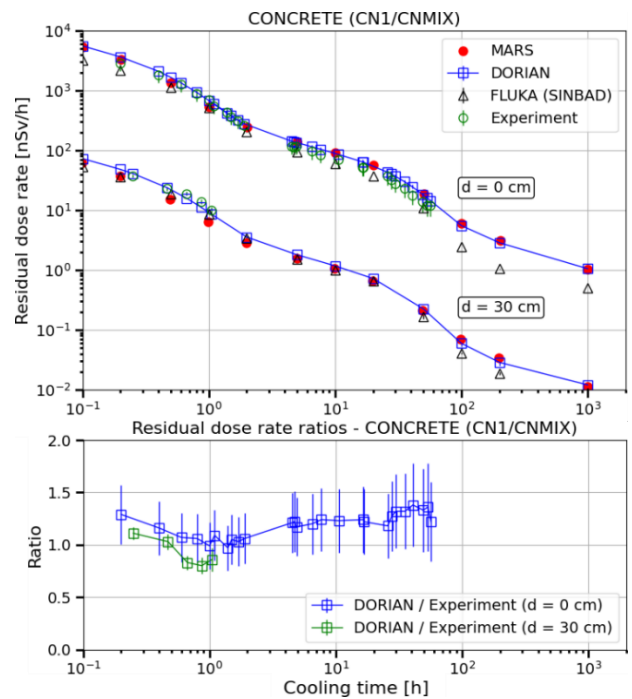

Figure 6: Comparison of residual dose rates for the Cn1 sample (03)

The residual dose rate results for the concrete sample are compared in Figure 6. Results from both MARS and DORIAN show good agreement with the experimental data, considering the uncertainties. At 30 cm, both MARS and DORIAN show an agreement with the experimental data within a range of ±20%. The dominant contributors to the dose rate for various cooling times can be observed as follows. For cooling times between 6 and 12 minutes, the dose rate primarily comes from the decays of $^{11}$C and $^{38}$K. For cooling times between 30 minutes and 1 hour the decay of $^{11}$C is the dominant contributor to the dose rate. For a cooling time of 2 hours, the dose rate is dominated by $^{24}$Na and $^{18}$F, whereas for cooling times between 5 and 50 hours, $^{24}$Na is the primary contributor. For cooling times from 100 to 200 hours $^{52}$Mn is the dominant contributor to the dose rate, while for a cooling time of 1000 hours, $^{22}$Na and $^{7}$Be are the primary contributors.

In Figure 2 through Figure 6, it is evident that for four of the five samples, the values measured at contact are lower than those provided by DORIAN. This implies that the samples might have been positioned slightly further from the detector, which is consistent with the understanding that the opposite scenario would be improbable.

*Impact of the simulation parameters on the results*

Further simulations were conducted to examine the effects of photonuclear interactions, muon pair production, and muon nuclear interactions on the results. Figure 7 shows the ratio of dose rates with and without the consideration of photonuclear and muon-associated processes. The influence of these processes varies across different samples, but generally, when these processes are considered, they can increase the dose rate by up to 20% for the considered samples and cooling times. The samples most impacted by these processes are copper, titanium, and concrete. Dose rates from the activated copper sample are predominantly affected at shorter cooling times, while those from the titanium sample are influenced both at short and more noticeably at longer cooling times.

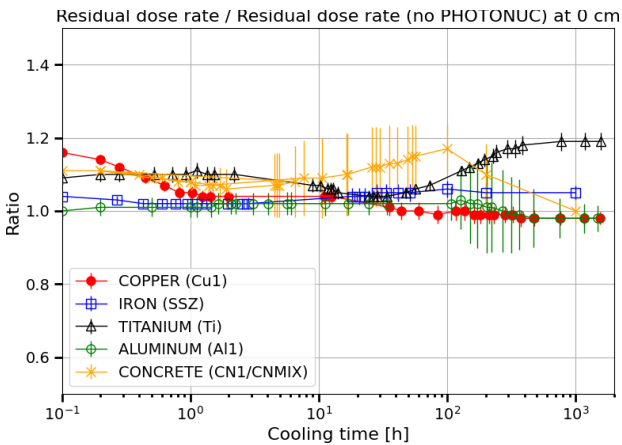

Figure 7: Influence of photonuclear and muon-related processes on dose rates. 'No PHOTONUC' denotes the scenario where photonuclear interactions, muon pair production, and muon nuclear interactions are excluded.

Lastly, we conducted further simulations using DORIAN, adopting the FLUKA simulation parameters presented in the SINBAD benchmark study. Figure 8 illustrates the ratio between our DORIAN results obtained using the experimental parameters and those derived from the FLUKA parameter set for each sample. Implementing the FLUKA parameters from the SINBAD benchmark appears to overestimate the residual dose rate from the copper sample by approximately 10% and from the aluminum sample by around 15%. However, for aluminum, this overestimation shifts to an underestimation after approximately 100 hours of cooling. Notably, for concrete samples, utilizing this parameter set appears to lead to an underestimation of the residual dose rate.

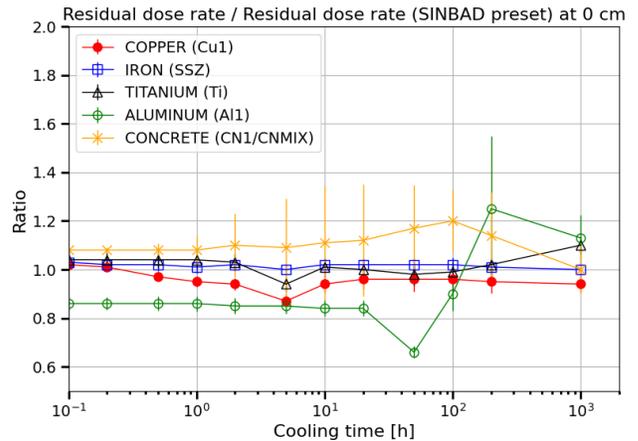

Figure 8: Comparison of residual dose rates using DORIAN results from experimental parameters versus those from the SINBAD benchmark's FLUKA parameter set.

## DORIAN SIMULATION RESULTS WITH MARS AND FLUKA ISOTOPE LISTS FOR COPPER TARGET

To investigate the overestimation of residual dose rates from the activated copper sample for cooling times less than 100 hours, as seen in Figure 1, we took the initial step of eliminating the potential influence of different flux-to-dose conversion factors. This was accomplished by creating isotope lists from both MARS and FLUKA for an irradiated copper target and using these lists as input for the Second Step calculations in a DORIAN analysis. By comparing the DORIAN results generated from the isotope lists produced by MARS and FLUKA, we aimed to determine the cause of the observed phenomenon.

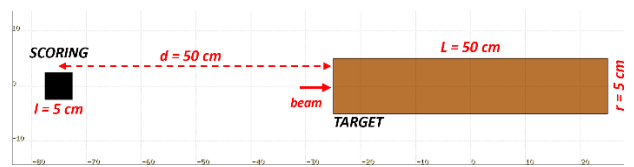

Figure 9: Simple geometry utilized in the DORIAN simulations using isotope lists from both MARS and FLUKA.

For this study, we opted for the simplistic setup detailed in the DORIAN paper [8]. This setup features a cylindrical copper target with a radius of 5 cm and a length of 50 cm, as shown in Figure 9. A proton beam with a momentum of 1 GeV/c impinges on the target at the center of its circular front face. After one year of irradiation at a beam intensity of one proton per second, the dose rate is measured 50 cm upstream from the impact face. Isotope lists for FLUKA and MARS were generated with 1.6e7 and 5.0e5 primary particles, respectively. We utilized the FLUKA preset labeled "PRECISIOn", with the EMF option activated. The particle transport threshold was set at 100 keV, except for neutrons, which were set at 1E-5 eV for both the irradiation and decay segments of the simulation. Full transport was applied to both light and heavy ions. Additionally, we activated the Coalescence mechanism, along with the Superposition model, within the energy range of 0.005 to 0.150 GeV/n. Two distinct second-step calculations were performed using DORIAN: the first evaluated dose-per-decay contributions from the decay of radionuclides recorded by MARS simulations, and the second assessed contributions from the decay of radionuclides recorded by FLUKA simulations. To enhance statistical accuracy, the decay of each individual isotope was replicated 30 times. In the final step of the DORIAN calculation, dose rates were separately computed based on both the MARS and the FLUKA isotope lists. This computation integrated contributions from the various radionuclides identified in the prior stage, considering cooling times that ranged from 10 minutes to 10 years.

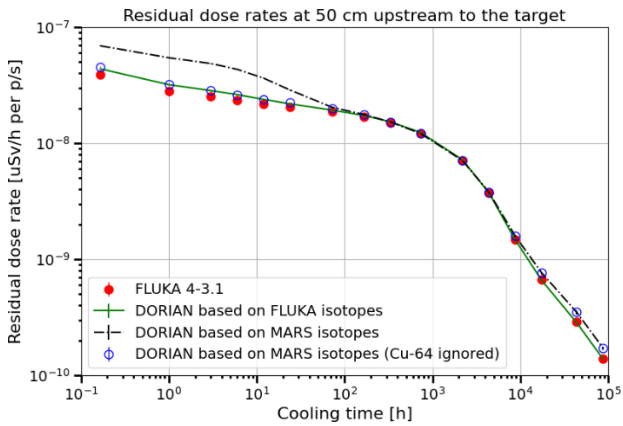

*Figure 10: Comparison of DORIAN results based on isotope lists from MARS and FLUKA.*

Figure 10 shows the values obtained from a one-step FLUKA simulation, as well as the DORIAN results derived from the isotope lists produced by both MARS and FLUKA. Figure 11 then presents the residual dose rate ratios between these different methodologies.

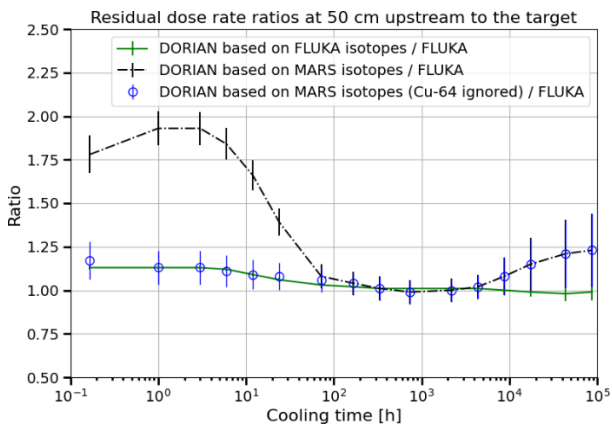

*Figure 11: Ratios of residual dose rates from DORIAN calculations using isotope lists from MARS and FLUKA.*

Two key observations emerge from the comparisons. Firstly, for short cooling times (<3 days) the residual dose rates calculated using the MARS isotope list are significantly higher (up to 90%) than those calculated based on the FLUKA isotope list. This observation aligns with the previously noted overestimation of dose rates in the case of the activated copper sample during cooling periods shorter than 100 hours, as shown in Figure 1. Furthermore, it is worth highlighting that in comparison to the one-step FLUKA simulation, DORIAN gives values up to 20% higher for short cooling times (<3 days), in agreement with the findings presented in the DORIAN paper [8].

To understand the cause of the elevated dose rates from MARS for short cooling times, we conducted an in-depth analysis. This involved comparing the dose rates of individual contributors. Specifically, we contrasted the dose rates of contributors from DORIAN results derived from both the MARS and FLUKA isotope lists for each considered cooling time. Table 2 presents the comparison for a cooling time of 1 hour, with contributors listed in descending order based on their contribution to the total, as calculated using the FLUKA isotope list. Given that these DORIAN calculations, regardless of whether the isotope lists from MARS or FLUKA are used, ensure the utilization of the same fluence-to-dose conversion factors, the results showcased in Table 2 can be interpreted as a comparison of production rates for isotopes significantly influencing the total residual dose rate. This comparison reveals a general agreement between MARS and FLUKA regarding the production rate of these isotopes within the copper sample. However, certain isotopes - namely $^{52}$Mn, $^{56}$Mn, and $^{62}$Zn - show differences by a factor of 2. Among these, $^{52}$Mn exhibits the highest contribution to the total dose rate, marginally under 7%. For the others, the contribution is less than 2%.

*Table 2: Comparison of residual dose rates from DORIAN calculations based on the MARS and FLUKA isotope lists for a 1-hour cooling time.*

| Contributor | | DORIAN(MARS) | | | DORIAN(FLUKA) | | | Ratio | |
|---|---|---|---|---|---|---|---|---|---|
| (Decay of isotope where @isotope were initially produced) | | Residual dose rate | Unc. | To total | Residual dose rate | Unc. | To total | MARS / FLUKA | Error |
| | | (uSv/h per pr) | | (%) | (uSv/h per pr) | | (%) | | (%) |
| **Total** | | 5.44E-08 | 3.07E-09 | | 3.19E-08 | 3.02E-10 | | 1.71 | 5.72 |
| Co-56 | @Co-56 | 6.31E-09 | 6.73E-10 | 11.60 | 7.52E-09 | 1.92E-10 | 23.57 | 0.84 | 10.96 |
| Cu-61 | @Cu-61 | 3.45E-09 | 1.10E-09 | 6.34 | 2.95E-09 | 7.73E-11 | 9.24 | 1.17 | 32.00 |
| V-48 | @V-48 | 2.67E-09 | 4.21E-10 | 4.91 | 2.63E-09 | 7.65E-11 | 8.25 | 1.02 | 16.02 |
| Co-58 | @Co-58, Co-58M | 4.95E-09 | 1.15E-09 | 9.10 | 4.70E-09 | 7.71E-11 | 14.71 | 1.06 | 23.23 |
| Co-58M | @Co-58M | - | - | - | 2.25E-09 | 8.32E-11 | 7.06 | - | - |
| Mn-52 | @Mn-52, Mn-52M | 4.06E-09 | 1.13E-09 | 7.45 | 2.15E-09 | 1.03E-10 | 6.73 | 1.89 | 28.21 |
| Mn-54 | @Mn-54 | 1.22E-09 | 2.63E-10 | 2.24 | 1.18E-09 | 3.29E-11 | 3.70 | 1.03 | 21.80 |
| **Cu-64** | **@Cu-64** | **2.25E-08** | **1.42E-09** | **41.33** | **1.06E-09** | **8.70E-12** | **3.32** | **21.21** | **6.36** |
| Cu-60 | @Cu-60 | 4.93E-10 | 1.19E-10 | 0.91 | 6.14E-10 | 3.41E-11 | 1.92 | 0.80 | 24.73 |
| Ni-57 | @Ni-57 | 6.16E-10 | 2.48E-10 | 1.13 | 5.41E-10 | 7.30E-11 | 1.70 | 1.14 | 42.46 |
| Co-55 | @Co-55 | 8.17E-10 | 3.72E-10 | 1.50 | 5.11E-10 | 4.67E-11 | 1.60 | 1.60 | 46.42 |
| Mn-56 | @Mn-56 | 8.74E-10 | 3.48E-10 | 1.60 | 4.48E-10 | 2.45E-11 | 1.40 | 1.95 | 40.23 |
| Sc-44M | @Sc-44M | - | - | - | 4.17E-10 | 2.88E-11 | 1.31 | - | - |
| Sc-44 | @Sc-44, Sc-44M | 7.38E-10 | 3.70E-10 | 1.36 | 6.91E-10 | 3.46E-11 | 2.16 | 1.07 | 50.35 |
| Zn-62 | @Zn-62 | 6.97E-10 | 3.91E-10 | 1.28 | 2.54E-10 | 2.97E-11 | 0.80 | 2.74 | 57.39 |
| Sc-46 | @Sc-46, Sc-46M | 7.17E-10 | 2.25E-10 | 1.32 | 5.21E-10 | 4.47E-11 | 1.63 | 1.38 | 32.52 |
| Co-60 | @Co-60, Co-60M | 6.22E-10 | 1.21E-10 | 1.14 | 5.00E-10 | 2.63E-11 | 1.57 | 1.24 | 20.23 |

However, a much higher disparity was observed in the production rate of $^{64}$Cu, as it was over 20 times higher in MARS than in FLUKA. Figure 12 shows the ratios of DORIAN results based on the MARS and FLUKA isotope lists for 1 hour of cooling time. The decay of $^{64}$Cu contributes approximately 3.3% to the total dose rate for the copper sample. This relatively low contribution, however, combined with the more than 20 times higher production rate in MARS, could lead to a significant increase in the total residual dose rate. In fact, the contribution of $^{64}$Cu to the total dose rate is about 41% when considering the DORIAN calculation based on the isotope list generated by MARS.

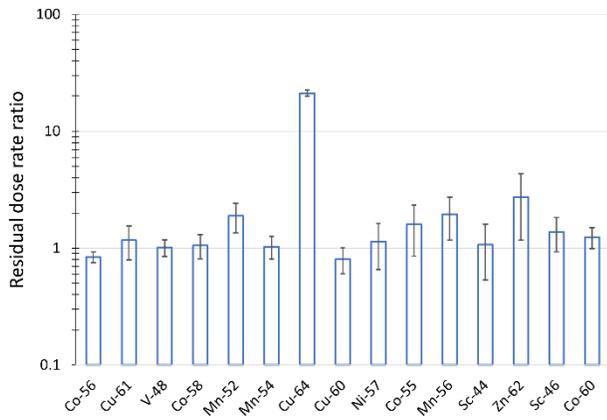

*Figure 12: Ratios of residual dose rates for dominant contributors from DORIAN calculations using the MARS and FLUKA isotope lists for a 1-hour cooling time.*

Given this disparity in production rates, $^{64}$Cu emerged as the prime contributor to the increased dose rates for short cooling times. To validate this hypothesis, we re-evaluated the DORIAN dose rate predictions using the isotope list from MARS in a modified third step process, deliberately excluding the contribution from $^{64}$Cu. As shown in Figure 10 and Figure 11 (represented by blue circles), this modification yielded a near-perfect alignment between the DORIAN results based on the MARS and FLUKA isotopes, with the only exception being for cooling times that extended beyond a year.

*Table 3: Comparing the contributions of $^{60}$Co decay to the residual dose rate in DORIAN results using MARS and FLUKA isotopes.*

| Contributor | Residual dose rate (uSv/h per pr/s) FLUKA | Contribution (%) | Residual dose rate (uSv/h per pr/s) MARS | Contribution (%) | Ratio (MARS/FLUKA) |
|---|---|---|---|---|---|
| Total | 1.38E-10 | | 1.72E-10 | | 1.25 |
| $^{60}$Co | 1.34E-10 | 97.2 | 1.67E-10 | 96.9 | 1.24 |
| $^{44}$Ti | 1.61E-12 | 1.2 | 3.54E-12 | 2.1 | 2.20 |
| $^{44}$Sc | 1.57E-12 | 1.1 | 1.35E-12 | 0.8 | 0.86 |
| $^{54}$Mn | 3.55E-13 | 0.3 | 3.65E-13 | 0.2 | 1.03 |

Statistics for cooling times longer than one year are relatively weak. However, the ratio graphs show that MARS results in higher dose rates, up to 25%. Table 3 shows the dose rates from the dominant contributors at a cooling time of 10 years. Considering that after a decade of cooling, 97% of the residual dose rate comes from the decay of $^{60}$Co, it is reasonable to conclude that MARS overestimates the production of $^{60}$Co isotopes in comparison to FLUKA.

*Table 4: Comparing the contributions of $^{60}$Co decay to the residual dose rate in DORIAN results using MARS and FLUKA isotopes.*

| Cooling time | Residual dose rate from 60Co (uSv/h per pr/s) | | Difference (%) |
|---|---|---|---|
| | FLUKA | MARS | MARS/FLUKA |
| 1 week | 4.98E-10 | 6.20E-10 | 24.38 |
| 2 weeks | 4.97E-10 | 6.18E-10 | 24.38 |
| 1 month | 4.94E-10 | 6.15E-10 | 24.38 |
| 3 months | 4.83E-10 | 6.01E-10 | 24.38 |
| 6 months | 4.68E-10 | 5.82E-10 | 24.38 |
| 1 year | 4.38E-10 | 5.45E-10 | 24.38 |
| 2 years | 3.84E-10 | 4.78E-10 | 24.38 |
| 5 years | 2.59E-10 | 3.22E-10 | 24.38 |
| 10 years | 1.34E-10 | 1.67E-10 | 24.38 |

As presented in Table 4, an approximate 25% increase in dose rate is noted across all cooling times. While this may align with the half-life considerations of $^{60}$Co, it is still noteworthy. Moreover, since the flux-to-dose conversion factor is the same using DORIAN regardless of the source of the isotope list, the observed difference in dose rate must come from the difference in the number of $^{60}$Co isotopes generated by MARS and FLUKA.

## CONCLUSION

Our comparative analysis of MARS and DORIAN simulations against experimental measurements underscores the accuracy and reliability of both models. Our investigation predominantly showcases strong agreement between DORIAN results and measurements, and similarly between MARS predictions and experimental data. Nonetheless, some minor discrepancies were found for both codes. Notably, for cooling times shorter than roughly 100 hours, MARS tends to overestimate the production of the $^{64}$Cu isotope in copper samples. This difference requires further investigation to clarify its underlying causes and implications.

## ACKNOWLEDGEMENTS

This work is supported by Fermi Research Alliance, LLC under contract No. DE-AC02-07CH11359 with the U.S. Department of Energy. This research used, in part, ALCC allocations at the Argonne Leadership Computing Facility (ALCF) and at the National Energy Research Scientific Computing Centre (NERSC) which are DOE Office of Science User Facilities supported under Contracts DE-AC02-06CH11357 and DE-AC02-05CH11231, respectively.